\documentclass{eas}
\usepackage{graphicx}
\usepackage{amssymb}
\begin{document}

\title{Pathways for observing stellar surfaces using 3D hydrodynamical simulations of evolved stars} 
\runningtitle{3D hydrodynamical simulations of evolved stars} 
\author{A.~Chiavassa}\address{Laboratoire Lagrange, Universit\'e C\^ote d'Azur, Observatoire de la C\^ote d'Azur, CNRS,
Blvd de l'Observatoire, CS 34229, 06304 Nice cedex 4, France}
\author{B.~Freytag}\address{Astronomical Observatory, Uppsala University, Regementsv{\"a}gen 1, Box 516, SE-75120 Uppsala, Sweden}
\begin{abstract}
Evolved stars are among the largest and brightest stars and they are ideal targets for the new generation of sensitive, high resolution instrumentation that provides spectrophotometric, interferometric, astrometric, and imaging observables. The interpretation of the complex stellar surface images requires numerical simulations of stellar convection that take into account multi-dimensional time-dependent radiation hydrodynamics with realistic input physics. We show how the evolved star simulations are obtained using the radiative hydrodynamics code CO5BOLD and how the accurate observables are computed with the post-processing radiative transfer code {\sc Optim3D}. The synergy between observations and theoretical work is supported by a proper and quantitative analysis using these simulations, and by strong constraints from the observational side.
\end{abstract}
\maketitle
\section{Introduction}

Evolved stars such as Red Supergiant (RSG) and Asymptotic Giant Branch (AGB) stars are characterized by low surface gravity (lower than $\sim\log g=1.0$) and their atmosphere is unstable against convection in deep layers with poorly defined boundaries. At effective temperatures lower than $\sim$ 4000K, spectra are so crowded with atomic and molecular spectral lines that there is little hope to see the continuum forming region, even at high spectral resolution. In addition to this, many molecular data, needed when a model atmosphere is computed or used for abundance analysis, is still not accurately known.
Ultimately, the derivation of their stellar parameters such as effective temperature and surface gravity is not trivial, and, in particular, the surface gravity is still highly uncertain. The understanding of the dynamical convective pattern of evolved stars is then crucial for the comprehnesion of the physics of these stars that contribute extensively to the chemical and dusty enrichment of galaxies.

\section{Modelling evolved stars}

We used realistic 3D radiative hydrodynamical (RHD) simulations of stellar convection (Table~\ref{simus}) computed with {\sc CO5BOLD} (Freytag {\em et al.\/} \cite{2012JCoPh.231..919F}), which solves the coupled equations of compressible hydrodynamics and non-local radiation transport. In the case of RSG and AGB stars, the \textsc{star-in-a-box} geometry is applied and the computational domain is a
cubic grid equidistant in all directions; the same
open boundary condition is employed for all sides of the computational box. The radiation transport for the simulations of evolved stars employs a short-characteristics
method, and, to account for the short radiative time scale, several
(typically 6 to 9) radiative sub$-$steps are performed per
global step. The RHD simulations employ a multi-group scheme where the frequencies that reach monochromatic optical depth unity within a certain depth range of the model atmosphere will be put into one frequency group with typically five wavelengths groups sorted according to the run of the monochromatic optical depth in a corresponding MARCS (Gustafsson {\em et al.\/} \cite{2008A&A...486..951G}) 1D model with a smooth transition to the Rosseland mean (OPAL opacities) in the optically thick regime. \\
Once the RHD simulation is relaxed, the snapshots are used for detailed post-processing treatment to extract interferometric, spectrophotometric, astrometric, and imaging observables are used to compare to the observations to tackle different astrophysical problems, as well as, to constrain the simulations. For this purpose, we use the 3D pure-LTE radiative transfer code {\sc Optim3D} (Chiavassa {\em et al.\/} \cite{2009A&A...506.1351C}) to compute synthetic spectra and intensity maps from
the snapshots of the RHD simulations. The code takes into account the
Doppler shifts due to convective motions. The radiative
transfer equation is solved monochromatically using pre-tabulated extinction coefficients as a function of temperature, density, and
wavelength. The lookup tables were computed using the same extensive atomic and molecular opacity data as the latest generation of
MARCS models. With the synergy between {\sc CO5BOLD} and {\sc Optim3D}, we produced a set of observables covering all the wavelengths from optical to far infrared. 

RHD simulations of evolved stars show a very heterogeneous surface caused by the peculiar granulation. Waves and shocks (with significant Mach numbers up to 5 or even larger) dominate in the outer layers together with the variation in opacity through the atmosphere. RHD models pulsate by themselves and do not have any dynamic boundary condition, the hydrodynamical equations
include the advection of momentum, which, after averaging over space and time, gives the dynamical pressure. RSG and AGB simulations are both characterized by large convective cells and strong shocks, however, AGBs have more extended and complex atmospheres with shocks pushing the mass much further out (Arroyo-Torres {\em et al.\/} \cite{2015A&A...575A..50A}) Typical RHD simulations of RSGs and AGBs have the stellar parameters reported in Table~\ref{simus}.\\

\begin{table}[!h]
\caption{Typical stellar parameters for RHD simulations of RSG (Chiavassa {\em et al.\/} \ \cite{2011A&A...535A..22C}) and AGB (Freytag {\em et al.\/} \ \cite{2008A&A...483..571F}) stars.}
\label{simus}  
\smallskip
\begin{center}
{\small
\begin{tabular}{cccccccc}
\hline
\noalign{\smallskip}

Simulation & Numerical & $M_{\mathrm{pot}}$ & $M_{\mathrm{env}}$ & $L$ &  $T_{\rm{eff}}$ & $R_{\star}$ &  $\log g$   \\
                  & resolution & $\!\!\!$[$M_\odot$]$\!$ & $\!\!\!$[$M_\odot$]$\!$ & $\!\!\!$[$L_\odot$]$\!$ & $\!\!\!$$[\rm{K}]$$\!$ & $\!\!\!$[$R_\odot$]$\!$ &  $\!\!\!$[cgs]$\!$  \\

\hline
RSG & $401^3$ &  12   & 3 &  90000 & 3500 & 840 & $-$0.33  \\
AGB & $401^3$ & 1 & 0.186 & 7000 & 2500 & 430 & $-$0.83  \\
\hline
\end{tabular}
}
\end{center}
\end{table}

\section{How do the RSG and AGB stars behave}

These simulations have been tested against observations with several techniques and used to interpret/predict the observations. \\
At low spectral resolution ($\lambda/\Delta\lambda\lessapprox20000$), the vigorous convective motions and inhomogeneities cause large fluctuations in the spectra that will affect {\sc Gaia} spectrophotometric measurements up to 0.28 mag in the blue photometric range and 0.15 mag in the red filter (Chiavassa {\em et al.\/} \cite{2011A&A...528A.120C}), as well as the stellar parameters, such as effective temperature, and photometric colors (Chiavassa and Freytag \cite{2014arXiv1410.3868C}). Stellar parameters determination resides on the dynamical effects of the atmosphere and are not constant with respect to temporal evolution. \\
At high spectral resolution ($\lambda/\Delta\lambda\gtrapprox70000$), atomic spectral line profiles reveal "C"-shaped bisector spanning values up to 5 km/s on a temporal scale of few weeks with good agreement with observations (Chiavassa and Freytag \cite{2014arXiv1410.3868C}). RHD simulations are also used to tune hydrostatic models that only use empirical calibrations, such as micro- and macro-turbulence velocities, to approximate the turbulent flow (Chiavassa {\em et al.\/} \cite{2011A&A...535A..22C}). \\
Eventually, interferometric techniques are crucial for evolved stars because allows, thanks to the high spatial and some spectral resolution, the direct detection and characterization of the convective pattern related to the surface dynamics. The two main observebles are: (i) the visibilities, which measure the brightness contrast of the source, are primarily used to determine the fundamental stellar parameters and the limb-darkening. (ii) The closure phases, which combine the phase information from three (or more) telescopes, provide direct information on the morphology of the observed object. Few examples of comparisons are from Chiavassa {\em et al.\/} (\cite{2010A&A...515A..12C}), who detected and measured the characteristic sizes of convective cells on the RSG star $\alpha$~Ori using visibility measurements in the infrared (Haubois {\em et al.\/} \cite{2009A&A...508..923H}). Moreover, they managed to explain the observation of $\alpha$~Ori from the optical to the infrared region using RHD simulations, showing that its surface is covered by a granulation pattern that, in the H and K bands, shows structures with small to medium scale granules (5-15 mas, while the size of the star at these wavelengths is $\sim$44 mas) and a large convective cell ($\sim$30 mas). These structures were later also observed by Montarg\`es {\em et al.\/} (\cite{2014A&A...572A..17M}, with AMBER instrument at VLTI) Montarg\`es {\em et al.\/} (submitted, and Montarg\`es's contribution in this conferece with PIONIER instrument at VLTI).

\section*{Questions}

\textit{O. De Marco:} How do you deal with the loss of information that you will have at the photosphere? How can you calculate the light if you do not know parameters such as temperature at the photosphere?\\
\textit{Answer:} We compute the emerging monochromatic intensity for each line-of-sight (i.e., column) of the 3D cube simulations. For every column we have a good-enough distribution of points in the temperature/density profiles from the outer layers towards the optical depth larger than one.\\

\textit{J. Groh:} I was wondering if you could compare the location of the "stellar radius" predicted by the stellar evolution models with those predicted by RHD simulations. Also, how far out of the simulations extend?\\
\textit{Answer:} RHD models don't actually predict the radius. It is rather a direct consequence of input parameters like luminosity and mass (and the efficiency of convection). Moreover, the radius predicted by evolutionary models is defined at about optical depth (e.g., Rosseland) equal to one. For evolved stars with effective temperature lower that about 4000K, there is a strong wavelength dependence of the stellar surface morphology with the optical region being almost completely hidden by electronic transitions of very abundant molecules. The near infrared region ($\sim$1.6$\mu$m), where H$^-$ continuum opacity encounter its minimum, is more suitable to see deeply in the photosphere where the flux forms (i.e., closer to optical depth equal to one). The RSG simulations's box is typically about 1.3 stellar radii large, while the AGB simulation ones is about 2.5 stellar radii large.\\

\textit{A. Lobel:} When you fit the Ti I line in $\alpha$ Ori, do you incorporate micro- and macro-turbulence velocities in the radiation transfer calculations? In an M-star such as $\alpha$ Ori micro-turbulence velocity from 1D models is small as $\sim$1 km/s. In A-type star however it can increase to 4-5 km/s. Astrophysical micro-turbulence is likely related to the connection zone. Do you think it would be possible to do 3D simulations of A-type stars as well to test if micro turbulence velocity is "not needed"?\\ 
\textit{Answer:} We do not use any micro- and macro-turbulence parameter in radiative transfer calculations. 3D simulations are expected to self-consistently and adequately account for non-thermal Doppler broadening of spectral lines. In Chiavassa {\em et al.\/} (\cite{2011A&A...535A..22C}), we calibrated the micro- macro-turbulence of 1D hydrostatic models with 3D ones and found values close to 1.45/1.28 km/s. For A type stars, old low-resolution 3D simulations fails to reproduce line profiles (see, e.g., Kochukhov {\em et al.\/} 2006, IAU Symposium, Xiv: 0610111). 



\begin{thebibliography}{99}
\bibitem[2015]{2015A&A...575A..50A} Arroyo-Torres, B., Wittkowski, M., Chiavassa, A., et al.\ 2015, A$\&$A, 575, A50 
\bibitem[2009]{2009A&A...506.1351C} Chiavassa, A., Plez, B., Josselin, E., \& Freytag, B.\ 2009, A$\&$A, 506, 1351 
\bibitem[2010]{2010A&A...515A..12C} Chiavassa, A., Haubois, X., Young, J.~S., et al.\ 2010b, A$\&$A, 515, A12 
\bibitem[2011a]{2011A&A...535A..22C} Chiavassa, A., Freytag, B., Masseron, T., \& Plez, B.\ 2011b, A$\&$A, 535, A22 
\bibitem[2011b]{2011A&A...528A.120C} Chiavassa, A., Pasquato, E., Jorissen, A., et al.\ 2011, A$\&$A, 528, A120 
\bibitem[2014]{2014arXiv1410.3868C} Chiavassa, A., \& Freytag, B.\ 2014, arXiv:1410.3868 
\bibitem[2008]{2008A&A...483..571F} Freytag, B., H{\"o}fner, S.\ 2008, A$\&$A, 483, 571 
\bibitem[2012]{2012JCoPh.231..919F} Freytag, B., Steffen, 
M., Ludwig, H.-G., et al.\ 2012, JCompPhys, 231, 919 
\bibitem[2008]{2008A&A...486..951G} Gustafsson, B., Edvardsson, B., Eriksson, K., et al.\ 2008, A$\&$A, 486, 951 
\bibitem[2009]{2009A&A...508..923H} Haubois, X., Perrin, G., Lacour, S., et al.\ 2009, A$\&$A, 508, 923 
\bibitem[2014]{2014A&A...572A..17M} Montarg{\`e}s, M., Kervella, P., Perrin, G., et al.\ 2014, A$\&$A, 572, A17 
\end{thebibliography}
\end{document}